\documentstyle[psfig,aps,preprint]{revtex}
\begin{document}
\draft
\pagestyle{plain}
\preprint{IMSC-98/12/57
}
\def\be{\begin{equation}}
\def\ee{\end{equation}}
\def\ba{\begin{array}}
\def\ea{\end{array}}
\def\bd{\begin{displaymath}}
\def\ed{\end{displaymath}}
\def\bea{\begin{eqnarray}}
\def\eea{\end{eqnarray}}
\def\bra{\langle}
\def\ket{\rangle}
\def\a{\alpha}
\def\b{\beta}
\def\g{\gamma}
\def\d{\delta}
\def\e{\epsilon}
\def\ve{\varepsilon}
\def\l{\lambda}
\def\m{\mu}
\def\n{\nu}
\def\G{\Gamma}
\def\D{\Delta}
\def\L{\Lambda}
\def\s{\sigma}
\def\p{\pi}
\title{
Constraints on mixing angles of Majorana  neutrinos
}
\author{Rathin Adhikari\thanks{
rathin@imsc.ernet.in}  and G. Rajasekaran\thanks{graj@imsc.ernet.in}}
\address{The Institute of Mathematical Sciences,
C.I.T Campus, Taramani,
Chennai-600 113, India
}
\maketitle
\date{26 November, 1998}
 
\begin{abstract}

By combining the inputs from the neutrinoless double beta decay and the
fits of cosmological models of dark matter with solar and atmospheric
neutrino data, we obtain constraints on two of the mixing angles of
Majorana neutrinos, which become stronger when coupled with the reactor
neutrino data.  These constraints are strong enough to rule out Majorana
neutrinos if the small angle solution of solar neutrino puzzle is borne
out. 

\end{abstract}
 

It is well known \cite{kay,kay2} that unless neutrinos are very massive
and nonrelativistic, or interact through both left and right-handed
currents, experimental data on neutrino-induced reactions cannot
distinguish between Dirac and Majorana neutrinos.  Neutrinoless double
beta decay remains as the only feasible tool to probe this question.
Although experiments \cite{telu,ger,baly} have so far provided only upper
limits on the rate of this decay, recent limits \cite{baly} combined with
other inputs on neutrino physics might already lead to important
information on whether neutrinos are Dirac or Majorana particles.  These
other inputs are indications from the analysis of the cosmic microwave
background for the presence of a hot dark matter component which is
presumably neutrinos of mass $\approx 1 $ eV \cite{primack1,silk,primack}
and the indications \cite{solar,Fukuda,lang} from the analysis of solar and
atmospheric neutrinos that neutrinos do oscillate and that the mass
differences among the three neutrinos are much smaller than this scale of
1 eV. 

We show in this note that if neutrinos are Majorana particles, a combined
study of all the above pieces of data leads to rather stringent
restrictions on two of the mixing angles that occur for three flavors of
neutrinos. If these results are then confronted with the values of these
mixing angles allowed by solar, atmospheric and reactor neutrino data, the
allowed regions are further narrowed and in fact, in a few cases, one is
already close to contradiction, thus leading to the conclusion that
neutrinos are not Majorana particles.

In the three flavor mixing scheme the neutrino flavor eigenstates
$\nu_{\alpha} = \nu_{e, \mu, \tau}$ are related to the mass eigenstates
$\nu_i = \nu_{1,2,3} $ by
\be
\nu_{\alpha} =    \sum_{i} U_{\a i} \n_i
\ee
\noindent
where $ U_{\a i}$ are the elements of the unitary mixing matrix $ U $.  We
note that for Majorana neutrino \cite{kay2,zuber} there are three $CP$
violating phases in contrast to the case of Dirac neutrino which has only
one phase. We use the parametrization \cite{zuber}: 
\begin{eqnarray}
U&=&\left(
\begin{array}{ccc} c_{\omega} c_{\phi}&s_{\omega} c_{\phi} e^{-i \d_1} 
& s_{\phi} e^{-i \d_2}
\\
-s_{\omega} c_{\psi} e^{i \d_1} -c_{\omega} s_{\psi} s_{\phi} e^{i 
\left(\d_2+\d_3\right)}  & c_{\omega} c_{\psi}  - s_{\omega}
s_{\psi} s_{\phi} e^{i \left(\d_3+\d_2-\d_1\right)}& s_{\psi} 
c_{\phi} e^{i \d_3}
\\
 s_{\omega} s_{\psi} e^{i \left( \d_1-\d_3  \right) }    
- c_{\omega} c_{\psi}    s_{\phi} 
 e^{i \d_2}    &        -c_{\omega} s_{\psi}
e^{-i \d_3}  -     s_{\omega}  c_{\psi}       s_{\phi} e^{i \left(\d_2 
-\d_1  \right)}
&   c_{\psi} c_{\phi}
\end{array}
\right).\nonumber\\
\end{eqnarray}
\noindent
where $ \d_1, \d_2$ and $\d_3 $ are the three $CP$-violating phases and $c
$ and $s$ stand for sine and cosine of the associated mixing angle
$\omega$, $\phi$ or $\psi$ placed as subscript. 

The rate for the neutrinoless double beta decay depends on the following
combination of the neutrino parameters \cite{kay2}: 

\bea
m_{0\nu\beta\beta}  =   \left|\sum_{i = 1, 2, 3 }  \eta_i \;{
U_{ei}}^2
\;m_{\nu_i}\right|  
\eea
\noindent
where $ m_{\nu_i} $ are the Majorana neutrino masses, $U_{ei}$ the
elements of the first row of the mixing matrix given in equation (2), $
\eta_i = \left( 1/i \right)  \eta_i^{CP} = \pm 1 $ and $\eta_i^{CP}$ is
the $CP$ parity of Majorana neutrino $\nu_i$.  Although neutrinoless
double beta decay has not yet been seen experimentally, the experimental
upper limits on this rate have recently improved to a significant extent. 
In particular one may refer to the results of the Tellurium \cite{telu}
and Germanium experiments \cite{ger,baly}.  The strongest upper limit so
far comes from the Germanium experiment \cite{baly,pri1} and it is
\bea
m_{0\nu\beta\beta}   &<& 0.56     \;\mbox{eV}    \;\;\;(  99  \%   \;\mbox{
confidence  level}  )  \nonumber  \\
&<& 0.46 \;\mbox{eV}    \;\;\;(  90 \% \;\mbox{confidence level}   )
\eea
These numbers have been obtained using the nuclear matrix elements
calculated in \cite{cal}.  We shall take into account the uncertainties in
this calculation (see below). Using the conservative number 0.56 eV we
have

\bea
 \left|\sum_{i = 1, 2, 3 }  \eta_i \;{  U_{ei}}^2
\;m_{\nu_i}\right|
 <     0.56   \;\;\mbox{eV}
\eea

Next we consider the fits to the recent data on the anisotropies of the
cosmic microwave background radiation \cite{micro} and the large scale
structure of the universe \cite{struc}.  Best fit \cite{primack,silk}
requires a mixture of $10 \%$ ordinary baryonic matter, $70 \%$ cold and
$20 \% $ hot dark matter with $\Omega_m = 1 $. If the hot component is
identified with neutrinos , the model implies \cite{primack1}
\be
\sum_{i= 1, 2,   3}   m_{\nu_i}   \approx 5  \;\mbox{eV}.  
\ee
Right hand side of (6) is not expected to be less than 3 eV for $\Omega_m
=1 $ \cite{pri2} as otherwise there is too much small scale power 
\cite{note1}. 
On the other hand, solar and atmospheric neutrino data suggest that the two
mass-squared differences among the three neutrinos are very small
\cite{solar,Fukuda,lang} :  $m_2^2 - m_1^2 \approx 10^{-5} {\mbox{eV}}^2$
or
smaller and $m_3^2 - m_1^2 \approx 10^{-3} - 10^{-2} {\mbox{eV}}^2$. 
Hence we take all three neutrinos as almost degenerate in mass and using
(6),
\be
m_{\nu_i}    \approx m_{\nu}  \approx  1.7\;  \mbox{eV}  
\ee
We shall allow $m_{\nu}$ to vary over a range around 1.7 eV.  This will
take care of the uncertainties of cosmological models as well as those of
the calculations of the nuclear matrix elements in double beta decay,
since only the ratio $0.56/m_{\nu}$ enters into our analysis. 
Any possible improvement  in the   neutrinoless double beta decay limit 
can also 
 be incorporated by scaling $m_{\nu}$ appropriately.

Combining all the inputs, we have the basic inequality
\bea
\left|  \left(\eta_1 \;\cos^2 \omega +       \eta_2 \;\sin^2 \omega
\;e^{-i 2 \d_1} 
\right) \cos^2 \phi \;  
+\eta_3 \;\sin^2    \phi    \; e^{-i 2 \d_2} 
 \right|   \;\;<  {0.56 \over m_{\nu}}
\eea
where $m_{\nu}$ is expressed in eV.  One can rewrite this inequality
 in terms  of 
two effective phases by combining $\eta_i$ and $\d_i$.   
However, to make our discussion on   $CP$ conserving  and   $CP$ violating
cases more transparent  we have kept both $\eta_i$ and  $\d_i$ above.  
We proceed to extract the bounds on
the mixing angles $\omega $ and $\phi $ implied by this inequality for
various choices of $\eta_i$ , $\d_i$ and typical values of $m_{\nu} $
favored by the cosmological models.  It is to be noted that in contrast to
the usual oscillation phenomena studied in neutrino physics, $CP$
violation plays an important role in neutrinoless double beta decay. 
 
We shall first consider $\d_1=\d_2= 0 $ in (8).  Out of eight possible
combinations for different values of $\eta_i$ in (8), four
combinations are equivalent to the other four, as only the overall
magnitude in the left hand side of this inequality is constrained.  So we
shall analyse (8) on the basis of four cases :  Case I :  $\eta_1 = \eta_2
= \eta_3 = \pm 1 $ ;  Case II : $\eta_1 = - \eta_2= \eta_3 = \pm 1$;  Case
III :  $- \eta_1 = \eta_2 = \eta_3 = \pm 1$ and Case IV: $\eta_1 = \eta_2 =
- \eta_3 = \pm 1$.  Case I is the natural choice for $\eta_i$, if there
exists a symmetry linking the three generations.  But then the left hand
side of (8) is unity and so unless $m_{\nu} \leq 0.56$ eV, the inequality
cannot be satisfied \cite{note2}. Since such low values of $m_{\nu}$ are
not expected
in the cosmological models, we conclude that the case of equal intrinsic
$CP$ parities for the three Majorana neutrinos is not favored.  The
allowed
regions in $\omega$ and $\phi$ for cases II, III and IV are plotted in
figures 1, 2 and 3 respectively for $m_{\nu} = 1.7$ eV. For both II and
III, small $\phi \;\;( \phi \approx 0 ) $ requires $\omega$ to be in the
region of $45^{ \circ}$.  Whereas small $\omega$ in II is not allowed at
all, small $\omega$ in III requires large $\phi$.  Case IV leads to a
constraint condition on $\phi$, that is independent of $\omega$ and
requires $\phi$ to be in the region of $45^{\circ} $.  Figures 
4 and 5  show the
total allowed regions for all possible combinations of $\eta_i$ (i.e., all
the cases I, II, III and IV) 
for $m_{\nu} = 6$ eV and 0.64 eV respectively.
One may note that for $m_{0 \nu \beta \beta}/m_{\nu}   \; \rightarrow 0$, the 
allowed values  of $\omega$ and $\phi$ are    constrained   to lie  on the 
three curves :
$\tan^2 \phi = \pm \cos 2 \omega, 1$. Our figure 4 (where 
$m_{0 \nu \beta \beta}/m_{\nu} = 0.093$)    gives a small 
width to these curves.

We next consider $CP$ violating case.  Now, the choice $\eta_1 = \eta_2 =
\eta_3 = \pm 1 $ , is \underline{not} ruled out. Thus $CP$ violation is
capable of changing the conclusions dramatically.  Figures 6-8 show the
allowed regions in $\omega$ and $\phi$ for a few choices of the parameters
$\d_1$ and $\d_2$. 

We have not considered ``maximal'' $CP$ violation $\delta_1 = \pi/2 $
and/or $\delta_2 = \pi/2$ since these cases are subsumed by the choices of
relative negative $\eta_i$, as far as our inequality (8)  is concerned. 
Therefore, it is important to keep in mind that our figures (1-5)  are
relevant for such ``maximal'' $CP$-violating cases also. In particular,
figure 1 and figure 2 can be considered to be the ``maximally'' $CP
$-violating case of $\delta_1 = \pi/2$ and $\delta_2 = 0$. 

Let us now compare the above bounds on the mixing angles with the results
of the analysis of data from reactor, solar and atmospheric neutrino
experiments. Before we make this comparison, we must      comment on the  
  role of $CP$ violation  in these experiments. We note that even for 
Majorana neutrinos, the oscillation phenomena are controlled 
  by a single $CP$-violating  phase     $\d$  \cite{doi}. 

The CHOOZ reactor experiment \cite{chooz} interpreted within
a three flavor framework \cite{graj} leads to the constraint
\be
\phi \leq  12.5^{\circ} 
\ee
  Although  $CP$ violation was neglected  in reference \cite{graj},   
it is easy to see that (9) is valid even if $CP$ is violated.  
Comparing (9) with figure 3, we see that case IV is ruled out, while
comparisons with figures 1 and 2 rule out all values of $\omega$ except
those in the region of $45^{\circ}$.  These are for $m_{\nu} =1.7 $ eV. 
The results are weakened for $m_{\nu} = 0.64 $ eV (figure 5)  and are
further strengthened for $m_{\nu} = 6 $ eV (figure 4).  If $CP$ is
violated, with $\d_1 = 0$ and $ \d_2 \neq 0 $ 
(figure 6),  the above conclusion is
still valid. But for $m_{\nu} =1.7$ eV, 
$\pi/8 \stackrel{\displaystyle
>}{\sim} \d_1 \stackrel{\displaystyle <}{\sim} 3 \pi/8$ and $\d_2 = 0$, (
figure 7) as well as  $\pi/8 \stackrel{\displaystyle
>}{\sim} \d_1, \; \delta_2 \stackrel{\displaystyle <}{\sim} 3 \pi/8$
(figure 8)
there is no allowed region at all after using (9).

It is important to note that as a consequence of (9), the 
effect of $CP$ violation in all the neutrino-oscillation phenomena
is much reduced since the $CP$ violating phase factors  $e^{\pm i \d}$ always 
occur in combination with $\sin \phi$, as $\sin \phi \; e^{\pm i \d}$. 
Hence it is legitimate to compare with the results of analyses of solar 
and atmospheric neutrinos  
  even though $CP$ violation is usually ignored  in those  
 analyses.

Analysis of solar neutrino data on the basis of the MSW effect
\cite{lisiraj} leads to the allowed $\omega $ values of either $2^{\circ}
- 3^{\circ}$ or $20^{\circ}-40^{\circ} $ for $\phi \leq 12.5^{\circ}$. So
comparing with the above result, we conclude that the
result of the present analysis contradicts the small $\omega $ MSW
solution ($\omega \approx 2^{\circ} - 3^{\circ}$ ), but there exists some
overlap with the large $\omega $ MSW solution ($\omega \approx 20^{\circ}-
40^{\circ}$). Also there is no contradiction with the vaccuum
oscillation as a solution of the solar neutrino problem since this also
requires large $\omega$ \cite{vigdel}. Finally there is no contradiction
with the results of the atmospheric neutrino analysis 
\cite{lisiraj} since
this does not involve $\omega$.

So far we have used the information on $m_{\nu}$ from cosmology to get
results on the mixing angles which were then compared with the results of
reactor, solar and atmospheric neutrinos. In view of the uncertainties of
cosmological models, one can ask what kind of information on the 
quasi-degenerate
mass for  Majorana neutrinos  can be obtained from
our analysis, if we drop the cosmological input completely. It is clear
from our figure 5, that for the small $\omega$ MSW solution of the solar
neutrino problem, we get an upper bound on $m_{\nu}$ of about 0.7 eV and
we have checked that this upper bound becomes 4 eV for the large $\omega$
MSW solution. 

The quantitative results  of our analysis are contained in  figures 1-8  
in the 
form of restrictions in $\omega$ and $\phi$. We may also state two  
qualitative conclusions  that emerge from our analysis.

1.  If neutrinos are Majorana fermions and $CP$ is conserved, all the
three neutrinos cannot have same $CP$ parities. This conclusion 
(which perhaps 
is well-known  and is included   here only for the sake of 
completeness) may be important for model building.

2. If neutrinos are Majorana fermions, the mixing angle $\omega $ cannot
be small. Hence, if the small $\omega $ solution turns out to be the only
correct solution of the solar neutrino problem, then neutrinos cannot be
Majorana fermions. This will have serious consequences  for models   
intended to explain small neutrino masses. 

The above results and conclusions are based on the present 
indications that the
experiments on the neutrinoless double beta decay require
$m_{0\nu\beta\beta} $ to be less than a fraction of an eV while models
with neutrinos as the hot component of dark matter require $m_{\nu} $ to
be higher than about 1 eV. The restrictions become more severe  
and the conclusions become stronger 
if the upper limit on $m_{0 \nu \beta \beta} $ decreases \cite{baudis11}   
and/or $m_{\nu} $ increases. 

After the first submission of our manuscript, we came across the papers of
Georgi and Glashow \cite{georgi} and of Branco, Rebelo and Silva-Marcos
\cite{branco} whose contents have partial overlap with our work. Our
analysis is more general than both these works in which $\phi$ is
put zero and further we have considered the   $CP$-violating cases in detail.

\hspace*{\fill}

\noindent
{\bf Acknowledgment}

We thank Raj Gandhi, Mohan Narayan and S. Uma Sankar for discussions and
Sandip Pakvasa for a useful communication.     We thank Rahul Sinha for an 
important discussion that clarified the role of $CP$ violation 
in neutrino oscillations.

\begin{figure}
\psfig{figure=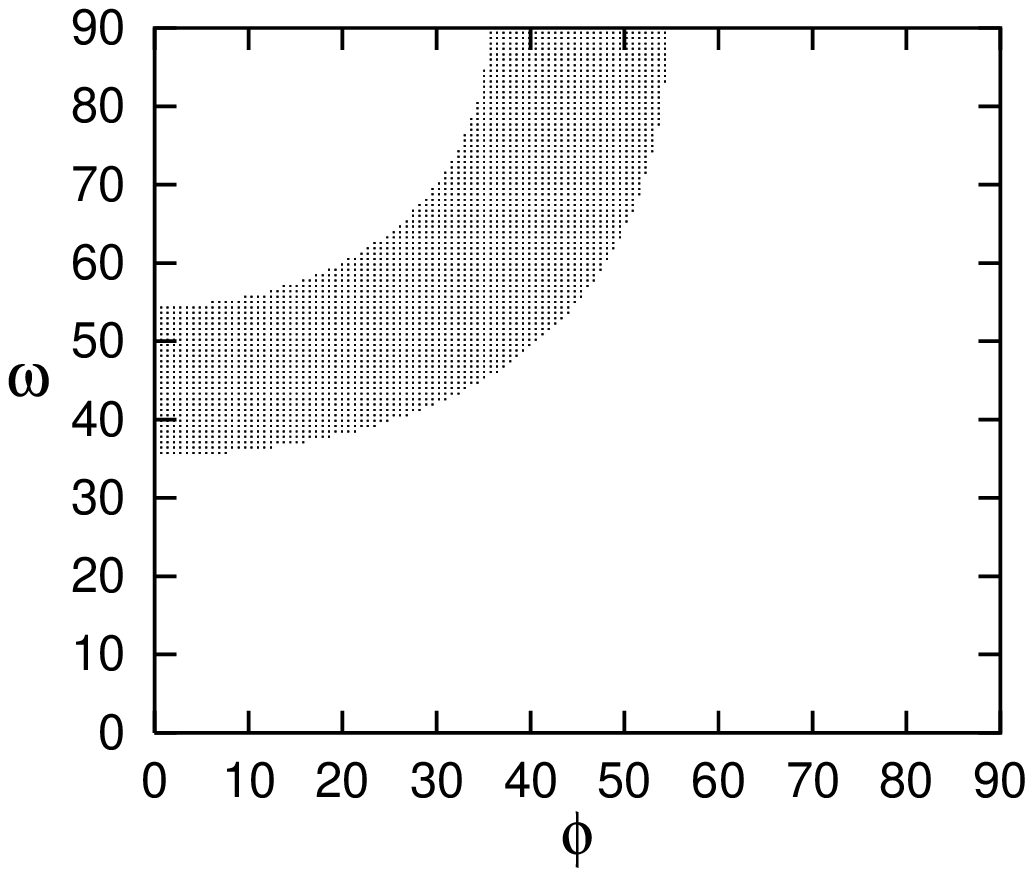,width=4.8cm,height=4.8cm} \vskip 0.00in
\caption{ The allowed region in $( \omega, \phi )$ is shown shaded for 
$m_{\nu} =1.7$ eV,
$ \eta_1 = -\eta_2 =\eta_3 = \pm 1  $ 
and $\d_1=\d_2 =0$.
}
\end{figure}
\vskip -0.1in
\begin{figure}
\psfig{figure=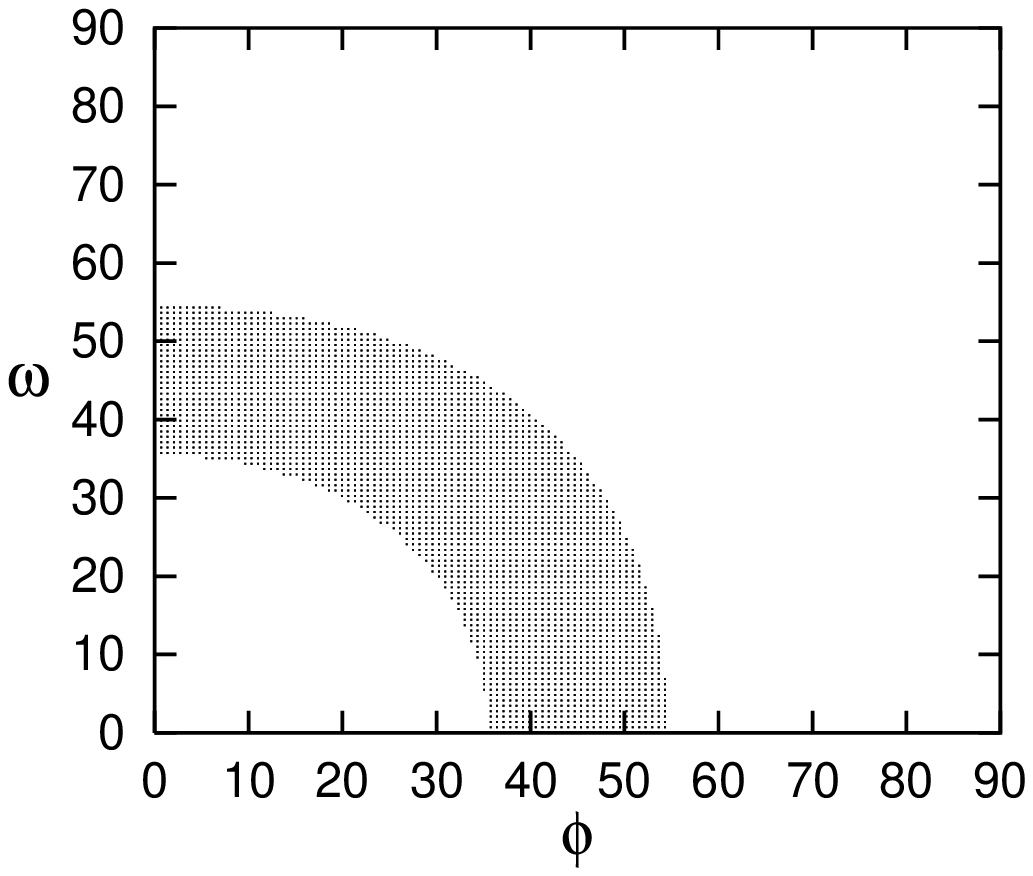,width=4.8cm,height=4.8cm}       \vskip 0.00in
\caption{ Same as figure 1 but
$ -\eta_1 =  \eta_2 =\eta_3 = \pm  1  $  .
}
\end{figure}
\vskip -0.1in
\begin{figure}
\psfig{figure=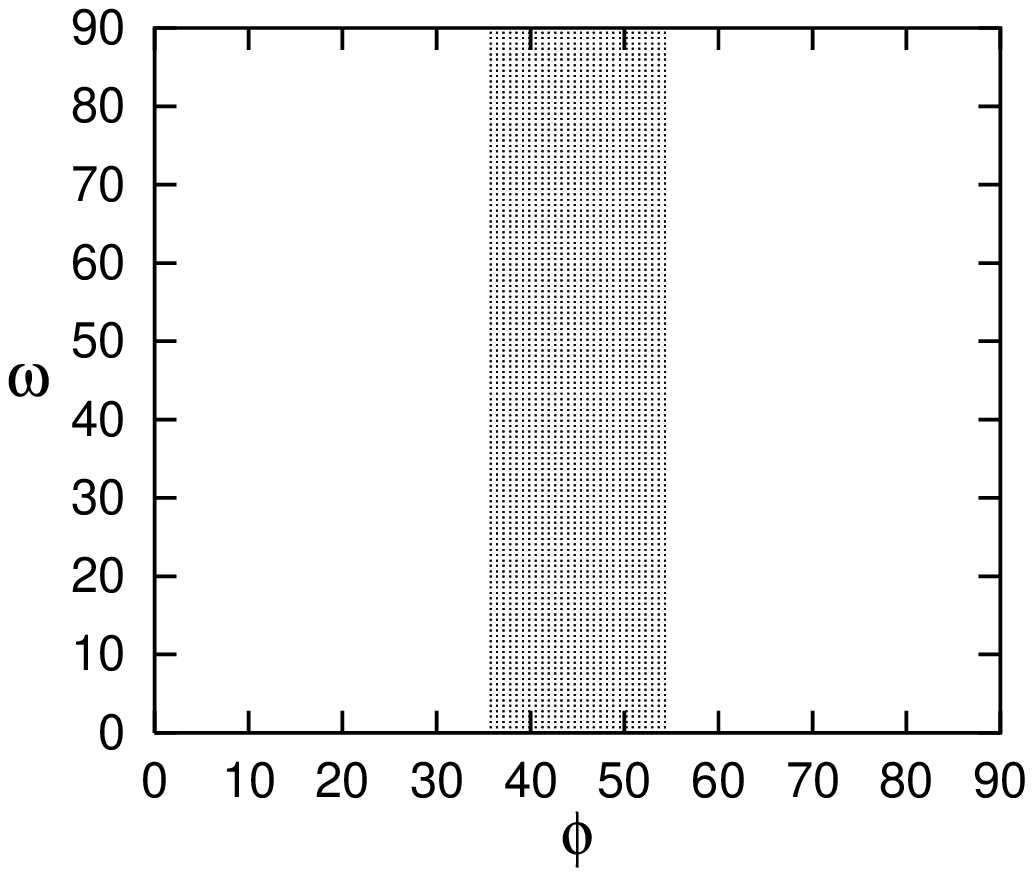,width=4.8cm,height=4.8cm} \vskip 0.00in
\caption{ Same as figure 1 but $\eta_1 =\eta_2 =-\eta_3 = \pm 1  $   .
}
\end{figure}
\vskip -0.1in
\begin{figure}
\psfig{figure=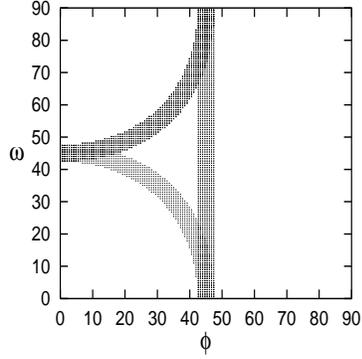,width=4.8cm,height=4.8cm} \vskip 0.00in
\caption{Allowed region for all possible combinations of
$\eta_i$,   $m_{\nu} =6$ eV,  and  $\d_1 = \d_2  = 0$.
} 
\end{figure}
\vskip -0.1in
\begin{figure}
\psfig{figure=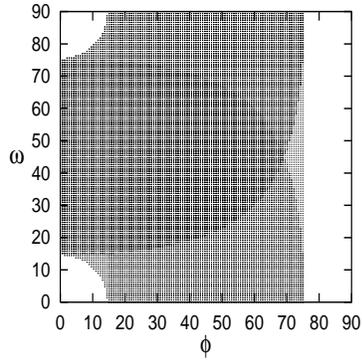,width=4.8cm,height=4.8cm} \vskip 0.00in
\caption{  
Same as figure 4 but for $m_{\nu} =0.64$ eV.} 
\end{figure}
\vskip -0.1in
\begin{figure}
\psfig{figure=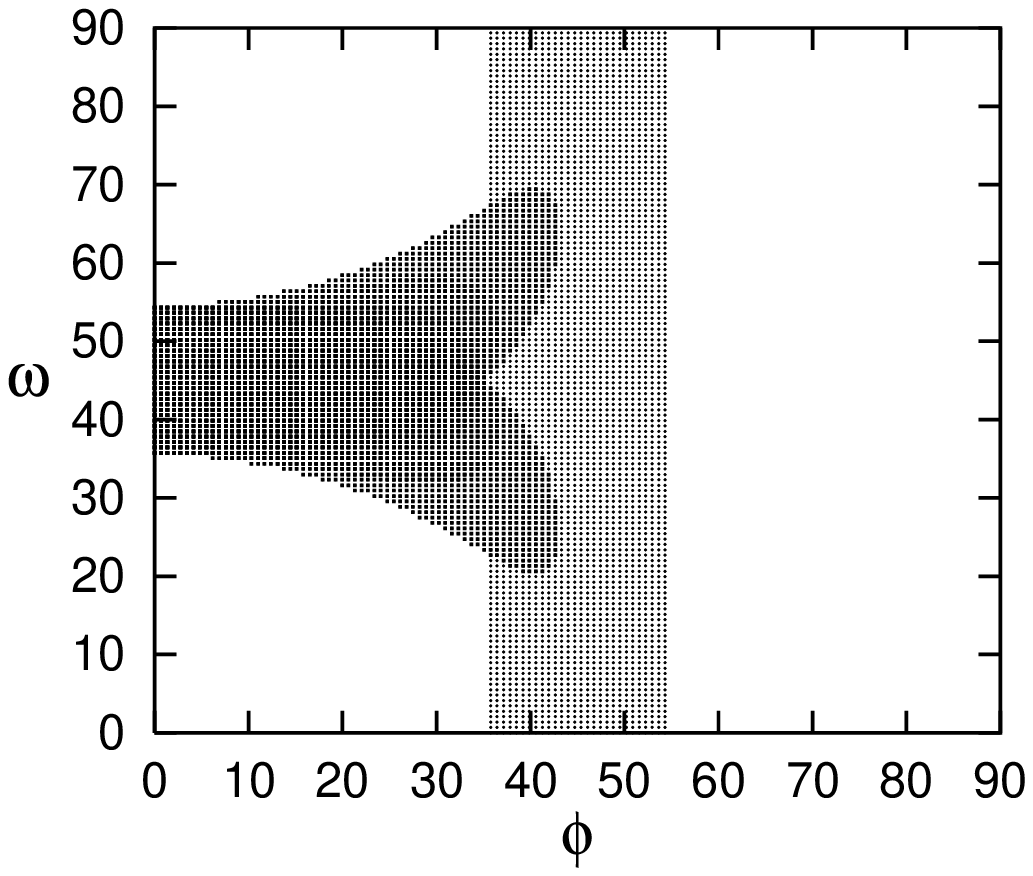,width=4.8cm,height=4.8cm} \vskip .10in
\caption{
Allowed region for all possible $\eta_i$, $m_{\nu}=1.7 $ eV   and $\d_1=0, \pi/8 \leq
\d_2 \leq 3 \pi/8$.
}
\end{figure}
\vskip -0.1in
\begin{figure}
\psfig{figure=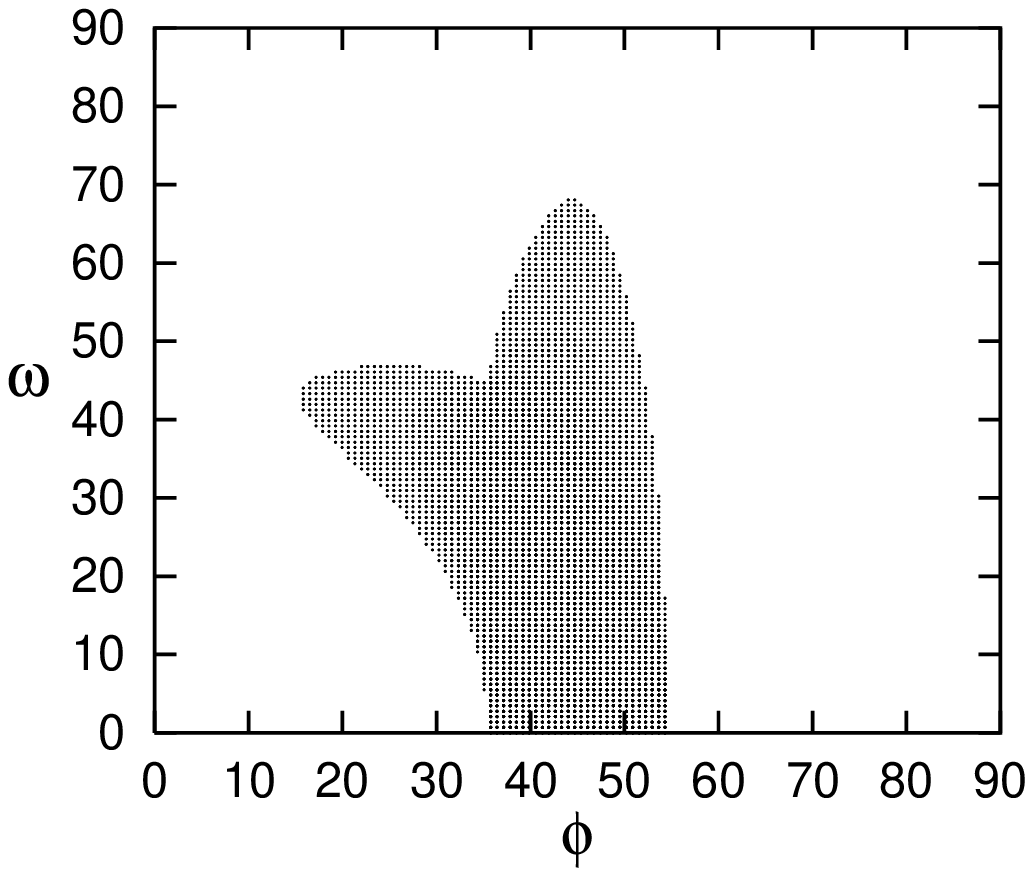,width=4.8cm,height=4.8cm} \vskip 0.00in
\caption{
Allowed region for all possible  $\eta_i$, $m_{\nu}=1.7 $ eV   
and $\pi/8 \leq \d_1 \leq
3 \pi/8,
 \d_2 = 0$.
}
\end{figure}
\vskip -0.1in 
\begin{figure}
\psfig{figure=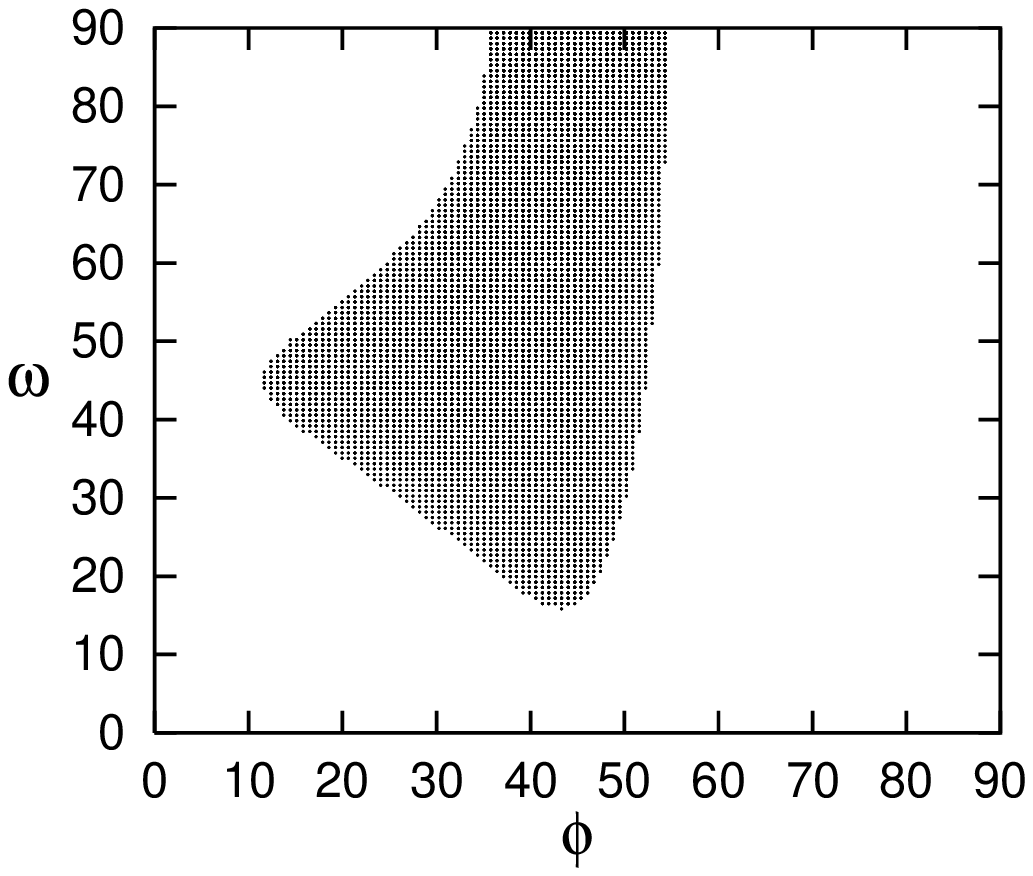,width=4.8cm,height=4.8cm} \vskip 0.00in
\caption{
Allowed region for all possible $\eta_i$,   $m_{\nu}=1.7 $ eV   
and $\pi/8 \leq \d_1,
\d_2 \leq 3 \pi/8$.
}
\end{figure}


\begin{thebibliography}{99}
\bibitem{kay}   B. Kayser, R. Shrock, Phys. Lett.   {\bf B 112}, 137 (1982).
 
\bibitem{kay2} R. N. Mohapatra and P. B. Pal, Massive neutrinos in Physics
and Astrophysics, World Scientific, (1991);  B. Kayser {\it et al}, The
Physics of massive neutrinos, World Scientific, (1989). 
 
\bibitem{telu} T.  Bernatowich {\it et al}, Phys. Rev. Lett.  {\bf 69},
2341 (1992);  Phys. Rev. {\bf C 47}, 806 (1993). 
 
\bibitem{ger} A. Balysh et al (Heidelberg- Moscow Experiment), Phys. 
Lett.  B {\bf 356}, 450 (1995). 

\bibitem{baly} M. G\"{u}nther {\it et al}, Phys. Rev. {\bf D 55}, 54
(1997); L.  Baudis {\it et al} (Heidelberg- Moscow Experiment), Phys. 
Lett.  B {\bf 407}, 219 (1997). 

\bibitem{primack1} J. R. Primack, Phys. Rev. Lett, {\bf 74}, 2160 (1995). 

\bibitem{pri2} J. R.  Primack, Private Communication. 
 
\bibitem{silk} E. Gawiser and J. Silk, Science, {\bf 280}, 1405 (1998); J.
R. Primack, Science, {\bf 280}, 1398 (1998). 

\bibitem{primack} J. R.  Primack , {\it Dark matter and Structure
Formation}, Presented in the Jerusalem Winter School 1996.
astro-ph/9707285; J. R. Primack and M. A. K. Gross astro-ph/9810204. 

\bibitem{solar} B. T. Cleveland et al, Nucl. Phys. {\bf B38} (Proc.
Suppl.) (1995) 47;  Y. Fukuda et al, Phys. Rev. Lett. {\bf 77 } , 1683
(1996) 

\bibitem{Fukuda} Y. Fukuda et al, hep-ex/9803006, 9805006, 9805021 and
9807003. 

\bibitem{lang} N. Hata and P. Langacker, Phys. Rev. {\bf D 56}, 6107
(1997); P. Langacker, hep-ph/9811460.

\bibitem{zuber} K. Zuber, Phys. Rep. {\bf 305}, 295 (1998); we 
have corrected some printing errors in the mixing matrix 
given in this report. 

\bibitem{pri1} H.  P\"{a}s, Private communication. 

\bibitem{cal} K. Muto, A. Staudt and H. V. Klapdor-Kleingrothaus,
Europhys. Lett. { \bf 13}, 31 (1990). 

\bibitem{micro} K. M. Gorski {\it et al} , Astrophys. J. Lett.  {\bf 430},
L89 (1994). 

\bibitem{struc} G. Efstathiou, J. R. Bond, S. D. M. White, Mon. Not. Roy. 
Astron. Soc.  {\bf 258}, P1 (1992). 

\bibitem{note1} Although the observations of high redshift supernovae   
 have been interpreted  \cite{highz}   in terms of a nonvanishing    
cosmological constant    $\Lambda$, cosmological models with    
$\Omega_m +  \Omega_{\Lambda}  = 1$  do not provide good fit to the 
structure data 
(CMBR anisotropies and the large-scale structure). Leaving this puzzle to 
be resolved by the cosmologists, we shall stick to equation (6) and 
deduce the   
consequences for neutrino physics.  

\bibitem{highz} S. Perlmutter  {\it et al}, Nature      {\bf 391}, 51 (1998);
P. M. Garnavich {\it et al} , ApJ  {\bf 493}, L53 (1998); A. G. Riess 
{\it et al},  astro-ph/9805201; B. P. Schmidt {\it et al}, astro-ph/9805200.

\bibitem{note2}  Although specific value   for   $m_{\nu}$ is mentioned
here as well as below,  it must be scaled suitably if the calculated
value  of the nuclear matrix element in      neutrinoless  
double beta decay changes or if the experimental limit  
on this decay improves, since only the
ratio  of the upper limit of   $m_{0 \nu \beta \beta}$ to  $m_{\nu}$ 
enters into our analysis.

\bibitem{doi} M. Doi {\it  et al}, Phys. Lett. {\bf B 102} , 323 (1981); S. M. 
 Bilenky {\it et al}, Phys. Lett.  {\bf B 94}, 495 (1980).

\bibitem{chooz} The CHOOZ Collaboration, M. Appolonio {\it et al } , Phys.
Lett. {\bf B 420}, 397 (1998) 

\bibitem{graj} Mohan Narayan, G. Rajasekaran and S. Uma Sankar, Phys. 
Rev. {\bf D 58}, 031301 (1998). 
 
\bibitem{lisiraj} M. Narayan {\it et al},
Phys. Rev. {\bf D 53}, 2809 (1996);  G. L. Fogli {\it et al}, 
Phys. Rev. {\bf D 54}, 2048 (1996); M. Narayan {\it et al}
, Phys. Rev. {\bf D 56}, 437 (1997)
; G. L. Fogli {\it et al},
Phys. Lett. {\bf B 434}, 333 (1998). 

\bibitem{baudis12} L. Baudis {\it et al}, hep-ex/9902014.

\bibitem{baudis11} In fact this has already happened; according to  reference 
\cite{baudis12},   $m_{0 \nu \beta \beta}    < 0.2 $ eV at 90 \%  
confidence level.

\bibitem{vigdel} S. L. Glashow {\it et al}, Phys. Lett. {\bf B
190}, 199 (1987); 
B. Faid {\it et al}, Phys.
Rev. {\bf D 55}, 1353 (1997); P. Osland {\it et al}, hep-ph/9806339; 
S. L. Glashow {\it et al},
hep-ph/9808470. 

\bibitem{georgi} H. Georgi and S. L. Glashow, hep-ph/9808293.

\bibitem{branco} G. C.
Branco {\it et al}, hep-ph/9810328.

\end{thebibliography}
\end{document}